# A Vortex Generator Flow Model Based on Self-Similarity

Clara Marika Velte[1]

[1]Department of Wind Energy, Technical University of Denmark, Niels Koppels Allé B-403, DK-2800 Kgs. Lyngby

## Nomenclature

| | | |
|---|---|---|
| $h$ | = | vortex generator height [m] |
| $L$ | = | vortex generator length [m] |
| $l$ | = | helical pitch [m] |
| $r$ | = | radial coordinate [m] |
| $u_z$ | = | axial (stream-wise) velocity component [ms$^{-1}$] |
| $u_\theta$ | = | azimuthal (rotational) velocity component [ms$^{-1}$] |
| $u_0$ | = | vortex convection velocity [ms$^{-1}$] |
| $U_\infty$ | = | free stream velocity [ms$^{-1}$] |
| $\varepsilon$ | = | vortex core radius [m] |
| $\delta$ | = | boundary layer thickness [m] |
| $\Gamma$ | = | circulation [m$^2$s$^{-1}$] |
| $\Theta$ | = | azimuthal (rotational) coordinate [°] |

## I. Introduction

VORTEX generators are devices that enhance mixing of the boundary layer and can thus transfer high momentum fluid closer to the wall and thereby control separation. They are commonly used on wind turbine blades and in aerospace, automotives and naval applications. They are also commonly used in mixing and heat transfer applications and the process industry. Many models for the generated vortices have been presented over the years. Theoretical models include, e.g., the one by Smith 1994 [1] and a 2-dimensional model presented by von Stillfried *et al.* [2] that is being developed and applied in computations to predict the generated vortices in an adverse pressure gradient. As for models incorporated into simulation codes most are variants of the BAY-model by Bender *et al.* [3], which introduces body forces using source terms in the Navier-Stokes equations to simulate the presence of a vane.

Velte *et al.* 2009 [4] showed that the vortices produced by vortex generators possess helical symmetry. This means, in effect, that the stream-wise ($u_z$, along the longitudinal vortex axis) and the rotational ($u_\theta$) flows are inter-related by a simple linear relation (in cylindrical coordinates)

$$u_z(r,\theta,z) - u_0(z) = -\frac{r}{l(\theta,z)} u_\theta(r,\theta,z) \qquad (1)$$

based on the helical shape of the vorticity lines [4,5], see Figure 1. $r$ is the radial coordinate with origin at the vortex center. There is a clear analogy to an electrical coil, where the current represents the vorticity and the induced magnetic field in the coil corresponds to the non-uniform induced velocity in $u_z$. The helical pitch corresponds to the spatial period of the vorticity lines. Applying (1) together with a Batchelor vortex model:

$$u_\theta(r,\theta,z) = \frac{\Gamma(z)}{2\pi r}\left[1-\exp\left(-\frac{r^2}{\varepsilon^2(\theta,z)}\right)\right]; \quad u_z(r,\theta,z) = u_0(z) - \frac{\Gamma(z)}{2\pi l(\theta,z)}\left[1-\exp\left(-\frac{r^2}{\varepsilon^2(\theta,z)}\right)\right] \qquad (2)$$

this complex flow can be described by four parameters: vortex core radius $\varepsilon(\theta,z)$, circulation $\Gamma(z)$, convection velocity $u_0(z)$ and helical pitch $l(\theta,z)$, leaving no restrictions on the shape of the vortex core.

The present work examines the downstream vortex evolution behind a cascade of vortex generators producing counter-rotating vortices in a boundary layer of negligible stream-wise pressure gradient. For the current study, which is an extension of the model presented in [4], these parameters are all seen to vary linearly in the downstream direction. Based on the experimental observations triggered by a previous study [6], the vortices generated by vortex generators have been observed to be self-similar for both the axial ($u_z$) and azimuthal ($u_\theta$) velocity profiles. The previous model, which is based merely on $u_z$ and $u_\theta$ at one single downstream location, can therefore be extended to include the downstream development of the vortex using self-similarity scaling arguments. This knowledge is important for fundamental understanding as well as for the aspect of applications, for which parametric experiments can be substantially reduced in terms of required time and cost by facilitating engineering models.

---

[1] Scientific Researcher, Department of Wind Energy, Division of Fluid Mechanics, Technical University of Denmark, Nils Koppels Allé, Building 403, 2800 Kgs. Lyngby. clara@alumni.chalmers.se

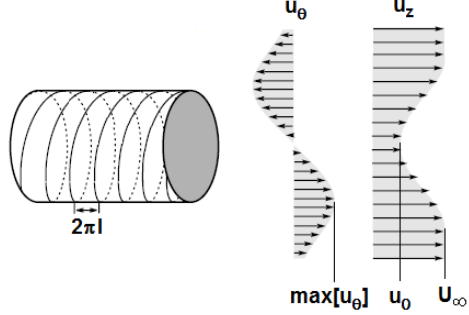

Fig. 1 Principal sketch of helical vortex and self-similarity normalization parameters.

## II. Experimental Method

The investigations were carried out in a low-speed closed circuit wind tunnel recording Stereoscopic Particle Image Velocimetry (SPIV) velocity fields in cross-planes at free stream speed $U_\infty$=1 m/s. The rectangular vortex generators, positioned directly on the flat test section wall, were the same height as the boundary layer thickness, $h=\delta$=25 mm, and with length $L=2h$=50 mm. The experimental setup is basically the same as in Velte et al. 2009 [4] with two exceptions: To minimize the influence of perturbing secondary structures [4,6], the object of study was a cascade of vortex generators mounted to produce counter-rotating vortices according to the optimal guidelines of Godard and Stanislas [7]. Further, the SPIV system was mounted on a traverse so that the equipment could be moved in the stream-wise direction to measure the development in several positions without requiring re-calibration. For each measurement position, 500 independent realizations were acquired. The disparity due to misalignment between the centre of the light sheet and the calibration target found in these investigations, typically around 0.05 pixels, was always smaller than the optimal measurement accuracy of the PIV system (~0.1 pixels). The linear dimensions of the interrogation areas (width = 1.55 mm, height = 1.04 mm) can be compared to the Taylor microscale and the Kolmogorov length scale estimated to $\lambda_f \approx$ 9 mm and $\eta \approx$ 0.5 mm from LDA measurements.

## III. Analytical Model

Self-similarity is based on the idea of self-preservation across scales. This kind of analysis is common for jets, but can conveniently be applied to wakes as well [8]. For the measured time-averaged far wake behavior it is posed that:

$$\frac{u_z(r,\theta,z) - u_0(z)}{U_\infty - u_0(z)} = fcn\left(\frac{r}{\varepsilon(\theta,z)}\right) \quad (3)$$

where $r$ is the radial coordinate, $u_z(r,\theta,z)$ the axial velocity, $U_\infty$ the free stream velocity and $\varepsilon(\theta,z)$ the vortex core radius (here taken as the characteristic width of the wake). Note that $\varepsilon=\varepsilon(\theta,z)$, $l=l(\theta,z)$ and $u_0=u_0(z)$ are all functions of $z$. The self-similarity relation (3) should also be compared to the velocity formulation (1), which has been confirmed to apply to the current flow [4], where the left-hand-side corresponds to the left-hand-side numerator in (3). Thus, it is natural to suspect that $u_\theta$ in the right-hand-side of (1) is also self-similar. A convenient scaling for the azimuthal velocity $u_\theta$ is to normalize it by the maximum value of $u_\theta$, see Figure 1. From self-similarity of $u_z$ and $u_\theta$, the model (2) can be extended to include the downstream development ($z$-dependence) of the vortices. Note that the vortex core radius was always chosen as the characteristic shear-layer width.

## IV. Description of Results

The averaged flow fields were recorded in cross-planes to give a full view of the vortex at each position. The measurements were processed in the same manner as in Velte *et al*. 2009 [4], applying a polar coordinate system $(r,\Theta)$ to the vortex with the origin at the vortex center for convenience and $\Theta$ measured from the horizontal line, see Figure 2. This made it simple to extract three-component velocity data diametrically along lines at any angle of choice. To cover the full rotational variations of the vortex, Figure 3 shows the axial ($u_z$) and azimuthal ($u_\Theta$) velocity profiles at four angles: $\Theta = 0°, 45°, 90°$ and $135°$. Each row represents one angle, as indicated in the figure. The first column shows $u_z$ and $u_\Theta$ for each downstream position, while the mid and right columns show $u_z$ and $u_\Theta$, respectively, with the applied self-similarity scaling. For some downstream positions, asymmetry occurs on the left side. This is mainly caused by the presence of the wall, perturbing the velocity field of the vortex core.

Figure 3 shows that applying similarity-scaling to both the $u_z$ and $u_\Theta$ profiles makes the curves collapse. The only exceptions are for some angles, showing distinct asymmetry, where the wall clearly influences the vortex core.

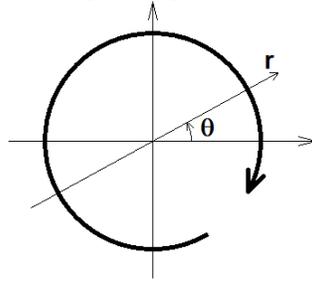

**Fig. 2 Polar coordinate system applied to the vortex.**

**Θ = 0°**

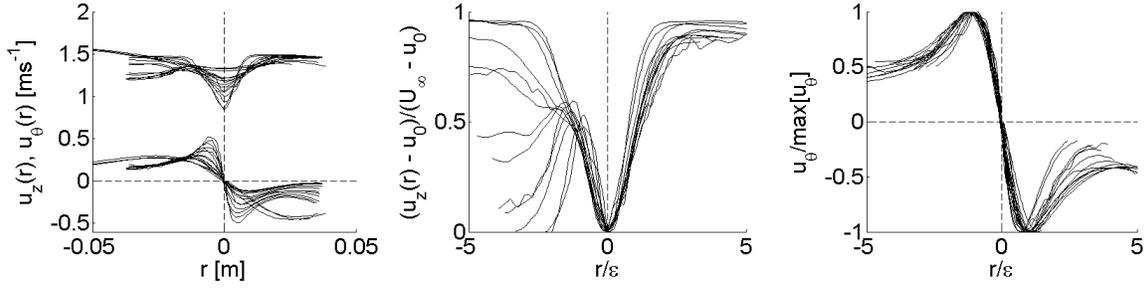

**Θ = 45°**

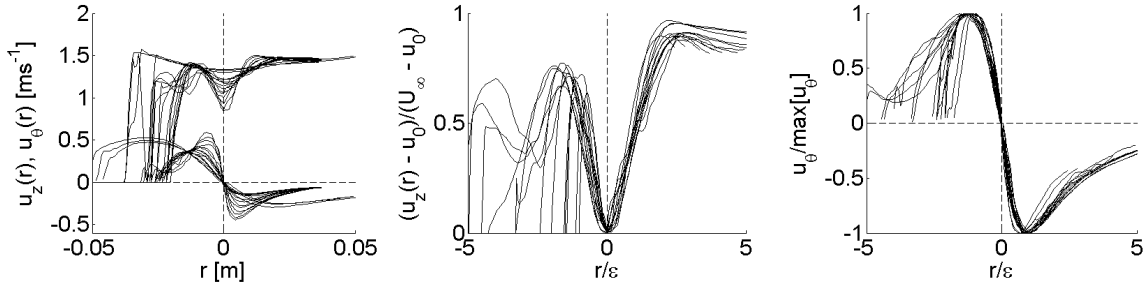

**Θ = 90°**

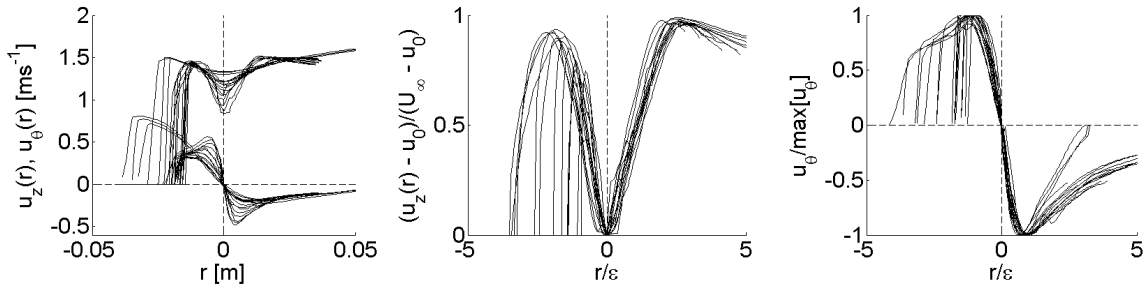

**Θ = 135°**

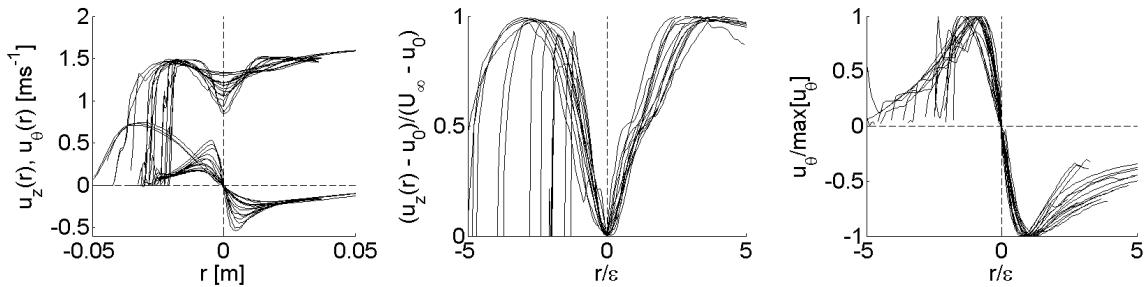

**Fig. 3  Velocity profiles for various angles and z/h = 2–13, showing the axial ($u_z$) and azimuthal ($u_\Theta$) profiles (left column) and the axial (middle column) and azimuthal (right column) scaled by self-similarity variables.**

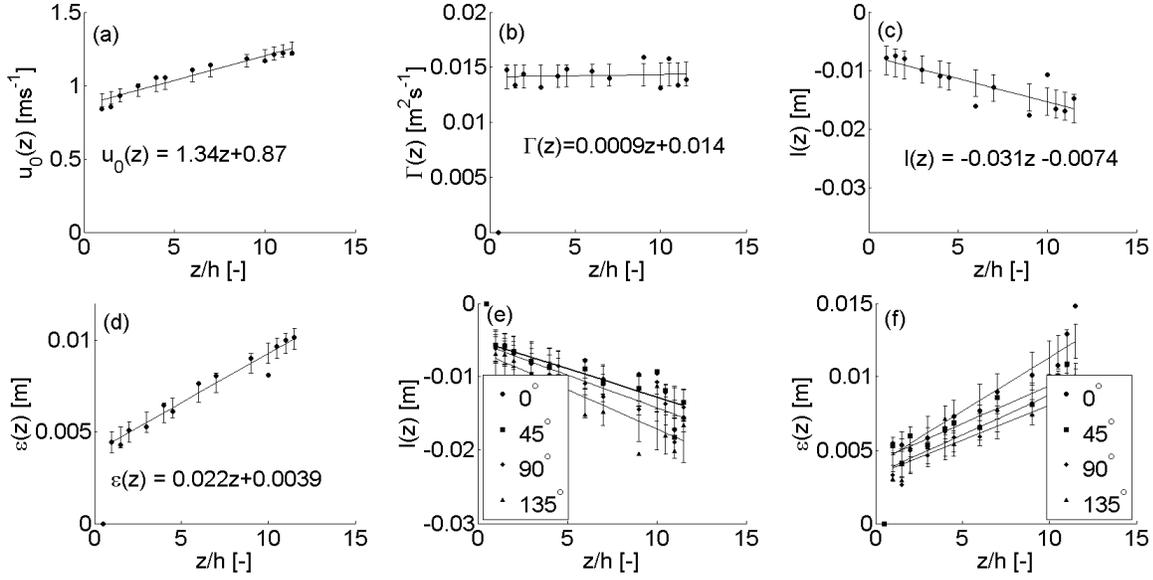

**Fig. 4 Downstream evolution of the characteristic vortex parameters in the stable wake.**

Figure 4 displays the stream-wise evolution of the helical parameters $u_0$, $\Gamma$, $l$ and $\varepsilon$, where the last two were averaged over $\Theta$ in 4(c,d) and shown for $\Theta=0°$, $45°$, $90°$ and $135°$ in 4(e,f). Up to about 1-2 vane heights downstream of the generator the wake is forming and therefore the model can't be applied there. Further downstream (at about 13 vane heights) perturbations in the secondary velocities due to the neighboring vortices become too large and the model can no longer be applied.

The velocity formulation (1) comprises, in addition to the axial and azimuthal velocity profiles, the convection velocity, $u_0$, and the helical pitch, $l$. Since both velocity profiles are self-similar, both $u_0$ and $l$ are expected to vary linearly along the downstream direction, which is confirmed by Figure 4. The negative sign of $l$ indicates that the helical vortex has a left-handed symmetry [5]. The coupling to $\varepsilon$ is far more complex (2), but the vortex core is expected to expand due to viscous diffusion (and stream-wise pressure gradients if present). The only parameter which is not expected to vary is the circulation, which should always be close to constant in a system of low viscous dissipation. Further, as can be seen from Figure 4(e) and (f), the variations in vortex core radius and pitch with rotational angle $\Theta$ are relatively small and can hence be neglected for present purposes.

## V.  Conclusion

The vortex generated by a rectangular vortex generator displays self-similar behavior for $z/h \approx 2\text{-}13$. This is observed for both the axial and azimuthal velocity profiles, which are linearly related. Consequently, all the characteristic parameters (convection velocity, helical pitch, circulation and radius) vary linearly along the downstream direction and the radius and pitch are only weakly dependent on the angle $\Theta$. Further, these linear trends in the helical parameters have been observed also for vortex development over adverse pressure gradients [9]. This means that, unless vortex breakdown occurs somewhere along the way [9,10], one should be able to interpolate the helical parameters linearly and thus model and describe the full vortex flow field. Further, the counter-rotating cascade arrangement of the vanes is seen to reduce the perturbations of neighboring secondary vortices [4,6,9].

## Acknowledgments

Martin O. L. Hansen and Sajjad Haider are gratefully acknowledged for helpful discussions. This work formed a portion of the Ph.D. dissertation of C.M.V, supported by the Danish Research Council, DSF, under Grant No. 2104-04-0020 and currently by EUDP-2009-II-Grant Journal No. 64009-0279, which are both gratefully acknowledged.